\documentclass[prl,superscriptaddress,twocolumn,showpacs,preprintnumbers,amsmath,amssymb]{revtex4}

\usepackage{graphicx}
\usepackage{dcolumn}
\usepackage{bm}
\usepackage{eucal}
\usepackage{ulem}
\usepackage{color}

\begin{document}

\title{Anomalous spin-waves and the commensurate-incommensurate magnetic phase transition in LiNiPO$_4$}
\author{T.~B.~S.~Jensen}
\affiliation{Materials Research Department, Ris\o{} DTU, National
Laboratory for Sustainable Energy, Technical University of
Denmark, DK-4000 Roskilde, Denmark}
\author{N.~B.~Christensen}
\affiliation{Materials Research Department, Ris\o{} DTU, National
Laboratory for Sustainable Energy, Technical University of
Denmark, DK-4000 Roskilde, Denmark} \affiliation{Laboratory for
Neutron Scattering, ETH Z\"{u}rich and Paul Scherrer Institut,
CH-5232 Villigen, Switzerland}
\affiliation{Nano Science Center, Niels Bohr Institute, University of Copenhagen,
Universitetsparken 5, 2100 K\o{}benhavn \O, Denmark}
\author{M.~Kenzelmann}
\affiliation{Laboratory for Neutron Scattering, ETH Z\"{u}rich and
Paul Scherrer Institut, CH-5232 Villigen, Switzerland}
\affiliation{Laboratory for Solid State Physics, ETH Z\"{u}rich,
CH-8093 Z\"{u}rich, Switzerland}
\author{H.~M.~R\o nnow}
\affiliation{Laboratory for Neutron Scattering, ETH
Z\"{u}rich and Paul Scherrer Institut, CH-5232 Villigen,
Switzerland}
\affiliation{Laboratory for Quantum Magnetism, \'{E}cole
Polytechnique F\'{e}d\'{e}rale de Lausanne (EPFL), 1015 Lausanne,
Switzerland}
\author{C.~Niedermayer}
\affiliation{Laboratory for Neutron Scattering, ETH Z\"{u}rich and
Paul Scherrer Institut, CH-5232 Villigen, Switzerland}
\author{N.~H.~Andersen}
\affiliation{Materials Research Department, Ris\o{} DTU, National
Laboratory for Sustainable Energy, Technical University of
Denmark, DK-4000 Roskilde, Denmark}
\author{K.~Lefmann}
\affiliation{Materials Research Department, Ris\o{} DTU, National
Laboratory for Sustainable Energy, Technical University of
Denmark, DK-4000 Roskilde, Denmark}
\affiliation{Nano Science Center, Niels Bohr Institute, University of Copenhagen,
Universitetsparken 5, 2100 K\o{}benhavn \O, Denmark}
\author{M.~Jim\'{e}nez-Ruiz}
\affiliation{Institut Laue Langevin, BP 156, F-38042 Grenoble
Cedex 9, France}
\author{F. Demmel}
\affiliation{Rutherford Appleton Laboratory, ISIS Facility, Didcot
OX11 0QX, Oxon, England}
\author{J.~Li}
\affiliation{Ames Laboratory and Department of Physics and
Astronomy, Iowa State University, Ames, Iowa 50011, USA}
\author{J.~L.~Zarestky }
\affiliation{Ames Laboratory and Department of Physics and
Astronomy, Iowa State University, Ames, Iowa 50011, USA}
\author{D.~Vaknin}
\affiliation{Ames Laboratory and Department of Physics and
Astronomy, Iowa State University, Ames, Iowa 50011, USA}

\date{\today}

\begin{abstract}
{Detailed spin-wave spectra of magneto-electric LiNiPO$_4$ have
been measured by neutron scattering at low temperatures in the
commensurate (C) antiferromagnetic (AF) phase with $T_N = 20.8$ K.
An anomalous low-energy mode is observed at the modulation vector
of the incommensurate (IC) AF phase appearing above $T_N$. A
linear spin-wave model based on Heisenberg exchange couplings and
single ion anisotropies accounts for all the observed spin-wave
dispersions and intensities. Along the $b$ axis an unusually
strong next-nearest-neighbor AF coupling competes with the
dominant nearest-neighbor AF exchange interaction and causes the
IC structure.}
\end{abstract}


\maketitle

{Strongly correlated systems with electronic and magnetic interactions have attracted much attention recently. Noted among them are the high-temperature superconductors, the colossal-magneto-resistance materials, and the multiferroics \cite{Cheong07}.  The latter group of materials possess  magnetic and ferroelectric phases that may even co-exist over a range of temperatures \cite{Kenzelmann05}.  For some, the correlations are manifested in the magneto-electric effect (ME) in which a magnetic field induces an electric polarization and an electric field induces a magnetic moment \cite{Eerenstein06}. The lithium orthophosphates Li$M$PO$_4$, with $M$ = Mn, Fe, Co and Ni constitute a prototypical iso-structural group of antiferromagnets (AF) that are ME below their N\'{e}el temperatures \cite{MercierThesis,Rivera1993}. Although they have been investigated for long \cite{MercierThesis}, a general consensus about the microscopic interactions leading to their ME effect has not yet been established \cite{Van-Aken2007,Jensen08}.

Here we focus on LiNiPO$_4$ which compared to its lithium orthophosphate (LO) counterparts have more complex magnetic ordering properties and a different temperature dependence of the linear ME coefficients \cite{MercierThesis,Vaknin2004}. LiNiPO$_4$ has an orthorhombic crystal structure with $Pnma$ (no. 62) symmetry, lattice parameters $a = 10.02$ {\AA}, $b = 5.86$ {\AA}, $c = 4.68$ {\AA}, and four formula units per unit cell \cite{Abrahams1993}. The magnetic Ni$^{2+}$ ions with spin $S = 1$ are located on the $4(c)$ sites and form buckled planes perpendicular to the crystallographic $a$ axis. All LO compounds have low temperature zero-field commensurate (C) AF phases, but only LiNiPO$_4$ passes via a first order phase transition at T$_N$ = 20.8 K into an incommensurate (IC) AF phase with ordering temperature T$_{IC}$ = 21.7 K. Above T$_{IC}$ short range correlations persist up to 40 K \cite{Vaknin2004}. In a recent study we have determined the magnetic structures and $(H,T)$ phase diagram as function of temperature and magnetic field up to 14.7 Tesla applied along the {\it c} axis \cite{Jensen08}. Here we also suggested a microscopic model accounting for the symmetry and temperature dependence of the measured ME coefficients \cite{Vaknin2004}. The present inelastic neutron scattering study is aimed at determining the magnetic couplings in LiNiPO$_4$ to improve our knowledge of the magnetic phases. Although the C phase has the general ordering vector ${\bf
k}_C$ = (0,0,0) we find an anomalous shallow minimum in the magnon
dispersion along the $b$ direction at a modulation vector ${\bf
q}_m \approx$ (0, 0.12, 0) (in reciprocal lattice units). This is in the range of
the IC ordering vector, ${\bf k}_{IC}$ = (0, $q$, 0), where $q$ is
increasing with temperature between $0.07 < q < 0.155$
\cite{Vaknin2004}. Analyzes of the spin-wave spectra show that the
minimum at ${\bf q}_m$ results from competing nearest-neighbor
(NN) and next-nearest-neighbor (NNN) AF interactions along the $b$
axis, and that the low temperature zero field AF structure is on the verge
between C- and IC-order.

\begin{figure*}[t]
\includegraphics[angle=-90,width=17.6cm]{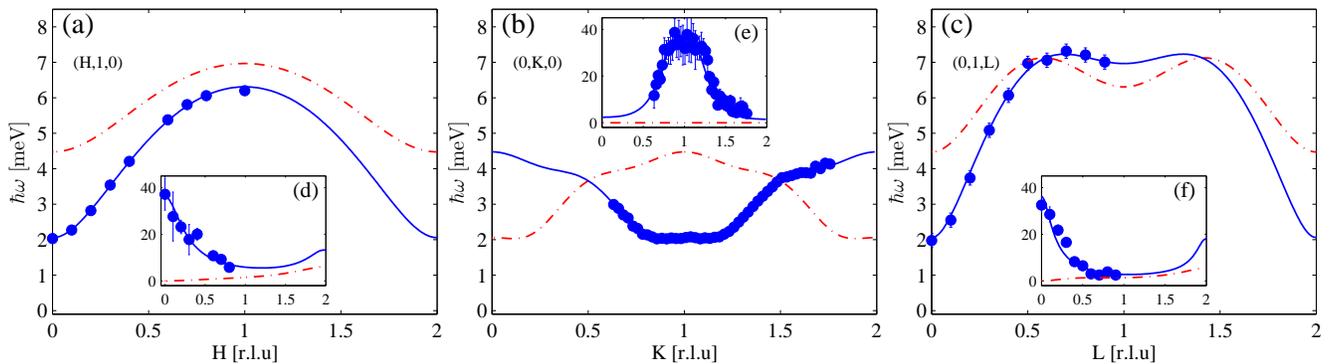}
\caption{Measured (filled circles) spin-wave dispersion along
three reciprocal directions, compared with a Holstein-Primakoff
spin-wave calculation (full and dotted lines) as explained in
the text. The insets show the corresponding measured and
calculated intensities (in arbitrary units) as function of wavevector. The dispersions along (H,1,0) and (0,K,0)
were measured at T = 2 K on RITA-II, while the dispersion along
(0,1,L) was measured at T = 10 K on HB1A.}
\label{fig:Figure1}
\end{figure*}

{A high quality single crystal with approximate dimensions $5 \times 5 \times 9$ mm$^3$ and weight 0.4 g was used for this study. Inelastic neutron scattering measurements were performed on the RITA-II spectrometer (SINQ, PSI) by varying the incident energy at constant $q$ to obtain dispersions along ($q$,1,0) and (0,$1+q$,0). The RITA-II uses a seven-blade analyzer and a final energy of 5 meV \cite{Bahl}. Spinwaves along the three principal directions, ($q$,1,0), (0,$1+q$,0) and (0,1,$q$) were measured on the HB1A spectrometer (HFIR, Oak Ridge) with 14.7 meV initial energy and were within uncertainties the same as those measured on RITA-II. More than an entire
Brillouin zone in the $bc$-plane was also measured on the IN8 spectrometer (ILL), using a 47 channel multi-analyzer-detector (MAD) \cite{Demmel} set for a final energy of 30 meV.}

\begin{figure}[b!]
\includegraphics[angle=-90,width=8.4cm]{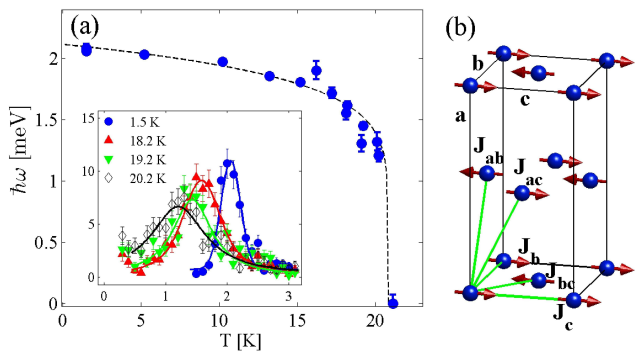}
\caption{(a) Spin-wave energy gap at $\mathbf{Q}=(0,1,0)$ as function
of temperature measured at RITA-II. The dashed line is a guide to
the eye. The inset shows constant $\mathbf{Q}=(0,1.055,0)$
scans at different temperatures with $\hbar\omega$ (in meV) along the
$x$-axis and measured intensity (in counts per minute) along $y$. (b) The magnetic
unit cell of LiNiPO$_4$ showing only
Ni$^{2+}$ ions. The depicted spin configuration and exchange
interactions are the ones used in the spin-wave model.} \label{fig:Figure2}
\end{figure}

The spin-wave dispersions and integrated intensities measured at RITA-II and HB1A along the $a$-, $b$- and $c$-directions at low temperatures are shown in Fig. \ref{fig:Figure1}. The ($q$,1,0) and (0,1,$q$) dispersions behave as expected for an antiferromagnet with an energy gap at the zone center, but the dispersion along (0,$1+q$,0) is unusual. It is almost a constant at low $q$ with a shallow minimum at $q \approx 0.12$, corresponding to the observed IC ordering vectors just below $T_{IC}$ \cite{Vaknin2004}. Generally, the (0,$1+q$,0) modes are lower in energy than along the two other principal
directions, and they soften the same way as the $q$ = 0 mode when the temperature increases towards the C-IC transition temperature $T_N$. Fig. \ref{fig:Figure2} shows the temperature dependence of the spin
wave gap at $\mathbf{Q}=(0,1,0)$. The inset displays spin-wave neutron scattering spectra at $\mathbf{Q}=(0,1.055,0)$, close to the $q$-value of the IC-modulation vector just above $T_N$.  No spin-waves were observed in the IC phase.

As a basis for our spin-wave model we adopt the AF ground state of the Ni$^{2+}$ ions determined by Santoro et al. \cite {Santoro1966} and Vaknin et al. \cite{Vaknin1999} with antiparallel spins pointing along the $c$-axis as shown in Fig. \ref{fig:Figure2}b. In fact, the spins are canted slightly away from the $c$ axis
\cite{Jensen08,JensenThesis2007}, consistent with a Dzyaloshinsky-Moria interaction, but this has negligible influence on the linear spin-wave model presented here. Accordingly, we use the following spin Hamiltonian for LiNiPO$_4$, with Heisenberg interactions $J_{ij}$ and standard single-ion anisotropy terms $D_{\alpha}(S^{\alpha})^2$ $(\alpha=a,b,c)$:
\begin{equation}
{\cal H} = \sum_{i,j} J_{ij}\textbf{S}_i\cdot\textbf{S}_j +
\sum_{i,\alpha}D_{\alpha}(S^{\alpha}_i)^2. \label{Hamiltonian}
\end{equation}
In the $bc$-plane the Ni-spins are coupled via super-exchange Ni$^{2+}$--O--Ni$^{2+}$ bonds, and our model includes NN interactions, $J_{bc}$, with a 3.81 {\AA} bond length, and two NNN, $J_b$ and $J_c$, with 4.71 {\AA} and 5.89 {\AA} bonds, respectively. For the couplings between the $bc$-planes we consider  NN interactions $J_{ab}$ and $J_{ac}$, which have 5.40 {\AA} and 5.50 {\AA} bond lengths (see Fig. \ref{fig:Figure2}b). The inter-plane couplings are mediated by PO$_4$-tetrahedra and may have a significant magnitude, e.g. as found in Li$_3$Fe$_2$(PO$_4$)$_3$ \cite{Zarestky2001}.

Using linear spin-wave theory we have calculated the spin-wave dispersions and intensities by a Holstein-Primakoff approach similar to the one described in \cite{Lindgaard1967}. The two spin-wave dispersions derived from Eq. (\ref{Hamiltonian})
are given by (for details see \cite{JensenThesis2007}),
\begin{equation}
\hbar\omega = \sqrt{A^2-(B \pm C)^2}, \label{spinwave}
\end{equation}
where,

\newpage

\begin{widetext}
\begin{equation*}
A\equiv 4S(J_{bc}+J_{ab})-2S[J_b(1-\cos(\textbf{Q}\cdot {\textbf{r}}_5))+J_c(1-\cos(\textbf{Q}\cdot {\textbf{r}}_6))+J_{ac}(2-\cos(\textbf{Q}\cdot {\textbf{r}}_7)-\cos(\textbf{Q}\cdot {\textbf{r}}_8))]+(S-1/2)(D_a + D_b),
\end{equation*}
\begin{equation*}
B \equiv (S-1/2)(D_a - D_b),\quad
C \equiv 2J_{bc}S[\cos(\textbf{Q}\cdot {\textbf{r}}_1)+\cos(\textbf{Q}\cdot {\textbf{r}}_2)]+2J_{ab}S[\cos(\textbf{Q}\cdot {\textbf{r}}_3)+
\cos(\textbf{Q}\cdot {\textbf{r}}_4)].
\end{equation*}
\end{widetext}
Here $\textbf{r}_i$ denote vectors between NN or NNN Ni-ions, and are:
$\textbf{r}_{1,2}=\tfrac12(\mathbf{b}\pm \mathbf{c})$, $\textbf{r}_{3,4}=\tfrac12(\mathbf{a}\pm \mathbf{b})$, $\textbf{r}_5=\mathbf{b}$, $\textbf{r}_6=\mathbf{c}$, $\textbf{r}_{7,8}=\tfrac12(\mathbf{a}\pm \mathbf{c})$.

The model parameters were determined by a simultaneous least squares refinement of the spin-wave dispersions in all three principal directions \cite{comment2}. The best fit is shown by full and dotted lines in Fig.\ \ref{fig:Figure1}, and the refined parameters are listed in Table \ref{tab:table1}. The couplings $J_{ac}$ and $J_c$
assure ferromagnetically ordered $ac$-planes that interact via $J_{bc}+J_{ab}$ to NN and $J_b$ to NNN $ac$-planes along the $b$-direction. All interactions  $J_{bc}$, $J_{ab}$ (NN) and $J_b$ (NNN) are AF which leads to frustration promoting the IC-ordering. We note that the $J_{ab}$ and $J_{ac}$ couplings in the $a$-direction are somewhat weaker than $J_{bc}$ and $J_b$. This is consistent with the findings in \cite{Vaknin1999} where magnetic short range fluctuations extending into the paramagnetic regime and a critical exponent $\beta=0.12$ of the C order parameter indicates a near-2D
magnetic ordering of the $bc$-planes. The single ion anisotropies, $D_a$ and $D_b$,
are both positive indicating that a $c$ axis magnetic moment is favored in the ground state, as observed experimentally. It should be noted that it is not possible from  Eq. (\ref{spinwave}) to determine which of the two is larger, $D_a$ or $D_b$.  However, comparing the calculated intensities to the measured intensities (insets of Fig. \ref{fig:Figure1}) determines unequivocally that $D_a < D_b$, as given in Table \ref{tab:table1}.

\begin{table}[b]
\caption{\label{tab:table1}The fitted spin coupling constants for
LiNiPO$_4$. \newline All units are in meV. By definition: $D_c
\equiv 0$ meV. }
\begin{ruledtabular}
\begin{tabular} {ccccccc}
$J_{bc}$&$J_b$&$J_c$&$J_{ab}$&$J_{ac}$&$D_a$&$D_b$\\
\hline
1.04(6)&0.670(9)&-0.05(6)&0.30(6)&-0.11(3)&0.339(2)&1.82(3)\\
\end{tabular}
\end{ruledtabular}
\end{table}

\begin{figure}[b]
\includegraphics[angle=-90,width=8.5cm]{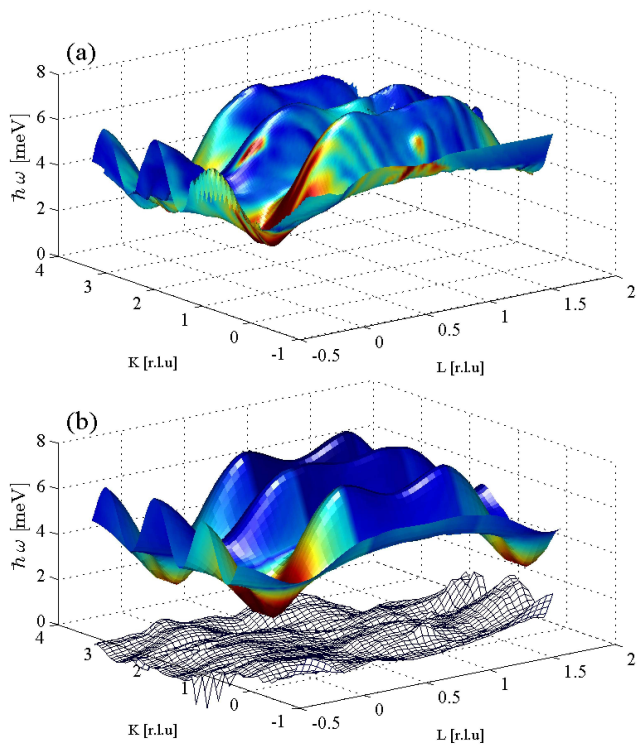}
\caption{Spin-wave dispersions in the $(0,K,L)$ plane corresponding to the most intense branch. The coloring is proportional to the observed spin-wave intensity (red for high intensities, blue for low). (a) Experimental data from IN8 at ILL. (b) Calculated data using the spin-wave model described in the text. The mesh around 0 meV shows the deviation in energy between (a) and (b).}
\label{fig:Figure3}
\end{figure}

\begin{figure*}[t]
\includegraphics[angle=-90,width=17cm]{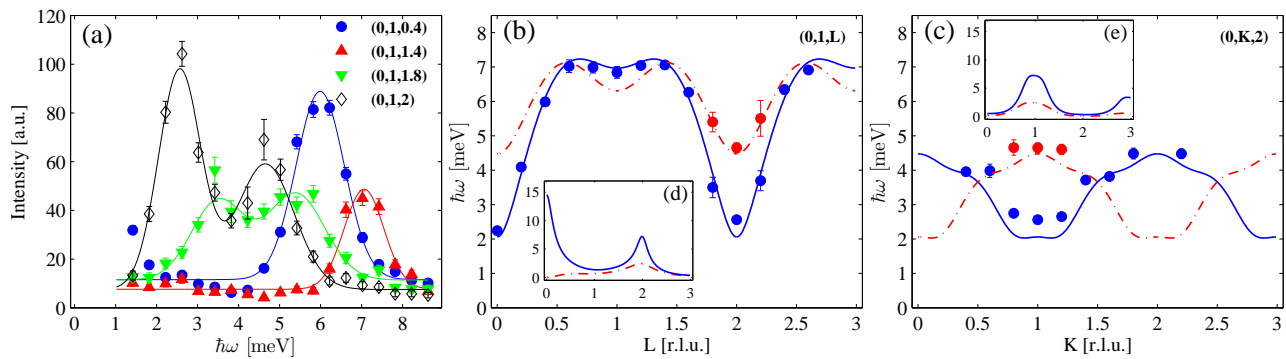}
\caption{(a) Constant $\mathbf{Q}=(0,1,L)$ cuts from IN8. At
$\mathbf{Q}=(0,1,0.4)$ and $(0,1,1.4)$ only a single branch is
visible, but at $\mathbf{Q}=(0,1,1.8)$ and $(0,1,2)$ both
branches are observed. (b) Dispersions along $(0,1,L)$
measured at IN8 (blue and red circles) compared to the calculated
spin-wave energies (full and dotted lines). The inset shows the
calculated intensity of the dispersions. Two branches are observed
at $\mathbf{Q}$ where the model predicts spin-waves with finite intensity
that are well separated in energy. (c)
Dispersions along $(0,K,2)$ measured at IN8 and compared to the
calculated spin-wave energies. The inset shows the calculated
intensity of the branches. Along $(0,K,0)$ (Fig.
\ref{fig:Figure1}b) the intensity of the second branch is
always zero, but here, along $(0,K,2)$, there are
$\mathbf{Q}$-values where both branches have intensity and can be
observed.} \label{fig:Figure4}
\end{figure*}

From the symmetry of the magnetic structure it can be shown that at least two non-degenerate dispersions are needed to account for the spin-wave
in the $b$-direction \cite{JensenThesis2007}. In our model calculations these two non-degenerate branches
is a result of the different anisotropies along $a$, $b$ and $c$. However, at RITA-II and HB1A (Fig.
\ref{fig:Figure1}) we observed only a single magnon dispersion, either because the
second branch had too low intensity (see Fig. \ref{fig:Figure1} insets), or because
the two dispersions were indistinguishable within instrumental resolution at the measured $\mathbf{Q}$-values.
Searching for the second branch and further confirmation of our model, we used the IN8 spectrometer to collect data from a larger range of $\mathbf{Q}$-values in the $bc$-plane. In Fig. \ref{fig:Figure3} we compare a 3D
color map of the most intense spin-wave dispersion measured at IN8 (a) to the results of our model calculation (b) using the interaction parameters of Table \ref{tab:table1}. The experiment at IN8 gave clear evidence of two non-degenerate magnon branches, as seen in Fig. \ref{fig:Figure4}. Here two dispersions were observed at scattering vectors where the calculations predicted that the branches were well separated in energy and both had finite intensity.

The exchange interactions established in this work may be used to explain the magnetic ground state. Maximizing $J(\mathbf{Q})=\sum J_{ij}e^{-i\mathbf{Q}\cdot\mathbf{R}_{ij}}$ at zero temperature shows that IC-order is favorable in a simple model of layered magnetic systems with competing interactions if the effective exchange interactions, $J_1$ and $J_2>0$, between NN and NNN ferromagnetic layers fulfill that $|J_1|<4J_2$ \cite{Nagamiya}. If this condition applies, the magnetic ordering vector $Q$ is determined by $\cos(Qd)=-J_1/4J_2$, where $d$ is the inter-layer distance between adjacent ferromagnetic layers. Since LiNiPO$_4$ has ferromagnetic layers perpendicular to the $b$-direction, we can test this condition. Here $J_1=2J_{bc}+2J_{ab}$ and $J_2=J_{b}$, while $d=b/2$. Using the exchange parameters in Table \ref{tab:table1} we find: $J_1 = 2.7(2)$ meV and $4J_2 = 2.68(4)$ meV, for the competing interactions. Within the uncertainties the magnetic ground state could therefore be either C or IC, and the system is close to an instability. A mean-field model using interaction parameters obtained from the spin-wave spectra predict a behavior qualitatively similar to that observed experimentally \cite{JensJensenPrivate}. Here it is found that the magnetic structure is IC just below T$_{IC}$, but that the extra lock-in anisotropy energy gained by the C structure becomes important at slightly lower temperatures and as result changes $q$ from $\mathbf{k}_{IC}$ to 0.

Jens Jensen is greatly acknowledged for illuminating discussions.
Work was supported by the Danish Natural Science Research Council
under DANSCATT, and by the Swiss NSF contract PP002-102831.
Experiments were performed at the SINQ neutron source at the Paul
Scherrer Institute, Institute Laue Langevin and HFIR. This manuscript has been authored,  
in whole or in part, under Contract No. DE-AC02-07CH11358 with the  
U.S. Department of Energy.


\begin{thebibliography}{01}
\bibitem{comment2} Since the energy gap does not change significantly between 2 and 10 K we used the same interaction parameters when fitting 2 K and 10 K.
\end{thebibliography}
\end{document}